\documentstyle[12pt,epsf]{ioplppt}

\begin{document}

\newcommand{\lesssim}{~\raisebox{.6ex}{$<$}\hspace{-9pt}\raisebox{-.6ex}
{$\sim $}~}

\sloppy

\jl{2}

\title{Two-photon detachment of electrons from halogen negative ions}[Two
photon detachment from halogens]

\author{G F Gribakin\dag , V K Ivanov\ddag ,  A V Korol\S\
and M Yu Kuchiev\dag\ftnote{4}{E-mails: gribakin@newt.phys.unsw.edu.au,
ivanov@tuexph.stu.neva.ru, Korol@rpro.ioffe.rssi.ru,
kmy@newt.phys.unsw.edu.au}}

\address{\dag School of Physics, The University of New South Wales,
Sydney 2052, Australia}

\address{\ddag Department of Experimental Physics, St Petersburg State 
Technical University, Polytekhnicheskaya 29, St Petersburg 195251,
Russia}

\address{\S Physics Department, Russian Maritime Technical
University, Leninskii prospect 101, St Petersburg 198262, Russia}


\begin{abstract}
Absolute two-photon detachment cross sections and photoelectron angular
distribution are calculated for halogen negative ions within lowest-order
perturbation theory. The Dyson equation method is used to obtain the
outer $np$ ground-state wave functions with proper asymptotic
behavior $P(r)\propto \exp (-\kappa r)$, corresponding to correct
(experimental) binding energies $E_b=\hbar ^2\kappa ^2/2m$. The latter
is crucial for obtaining correct absolute values of the multiphoton
cross sections (Gribakin and Kuchiev 1997 {\em Phys. Rev. A} {\bf 55} 3760).
Comparisons with previous calculations and experimental data are performed. 

\end{abstract}




\maketitle

\section{Introduction}\label{Intr}

Halogen negative ions have been subject of experimental and theoretical
multiphoton detachment studies for over thirty years (Hall \etal 1965,
Robinson and Geltman 1967). Apart from the negative hydrogen ion which
traditionally receives a lot of attention, especially from theorists,
they are definitely the most studied negative ions. Nevertheless, there
are very few firmly established results on the absolute values of the cross
sections and photoelectron angular distributions in multiphoton processes.
A number of experimental works reports the cross sections and angular
asymmetry parameters measured in two-photon detachment at selected photon
energies in F$^-$ and Cl$^-$ (Trainham \etal 1987, Blondel \etal 1989, 1992,
Kwon \etal 1989, Davidson \etal 1992, Sturrus \etal 1992, Blondel and
Delsart 1993). For heavier halogen ions, Br$^-$ and I$^-$, the experimental
data are scarce (Hall \etal 1965, Blondel \etal 1992, Blondel and
Delsart 1993).

On the theoretical side, there were pioneering two-photon detachment
calculations by Robinson and Geltman (1967) performed using a model
potential approach, and a number of other perturbation-theory calculations
employing the Hartree-Fock (HF) approximation for the ionic ground state, and
either plane or HF waves for the photoelectron in the continuum
(Crance 1987a,b, 1988). The latter were applied to study $n$-photon
detachment cross sections and photoelectron angular distributions from
halogen negative ions for $n$ up to five. Jiang and Starace (1988) used a
transition-matrix approach and examined the contribution of the lowest-order
correlation processes in the two-photon detachment from Cl$^-$. They
showed that the role of correlations is small just above the threshold but
increases with the photon energy and reaches about 20\%,
compared with the HF result. Later on Pan \etal (1990) and Pan and Starace
(1991) performed similar calculations of the two-photon detachment
cross section and angular distribution for F$^-$. Using first-order
perturbation theory in electron interaction they included more correlation
corrections and estimated the contribution of the many-electron effects
at 10 to 20\% in the cross sections, but almost negligible in the angular
distribution asymmetry parameters. More recently van der Hart (1996) used
the $R$-matrix Floquet theory to calculate multiphoton detachment from
F$^-$ and Cl$^-$. He found a discrepancy with the results of the
transition-matrix approach in F$^-$ and Cl$^-$ within about 30\%.

In the above works, except that of Robinson and Geltman, the halogen ion
ground state was described either in the HF, or in a few-state configuration
interaction approximation. Consequently, the binding energy of the outer
electron, as obtained from the calculation, was never in good agreement with
the experimental electron affinity, and experimental energy values 
were used in the calculations of the multiphoton amplitudes and cross sections.
On the other hand, the asymptotic behaviour of the ground-state wave function
in these calculations remained incorrect. This may seem to have introduced
only a small error in the calculation, since the bound-state wave function
in the asymptotic region is small. However, as shown in the adiabatic
theory of multiphoton detachment from negative ions (Gribakin and Kuchiev
1997a, 1997b), the asymptotic behaviour of the bound-state wave function
is crucial for obtaining correct absolute values of the probabilities of
multiphoton processes. This theory based on the Keldysh approach (Keldysh 1964)
approach shows that the electron escape from the atomic system in a
low-frequency laser field takes place at large distances,
\begin{equation}\label{large}
r\sim 1/\sqrt{\omega }\sim \sqrt{2n}/\kappa \gg 1,
\end{equation}
where $\omega $ is the photon frequency, $\kappa $ is related to the
initial bound state energy, $E_0=-\kappa ^2/2$, and $n$ is the number of quanta
absorbed (atomic units are used throughout). Accordingly, the multiphoton
detachment rates are basically determined by the long-range asymptotic
behaviour of the bound-state wave function, namely by parameters $A$ and
$\kappa $ of the corresponding radial wave function
$R(r)\simeq Ar^{-1}e^{-\kappa r}$. This result is obtained using
the length form of interaction with the laser field, which proves to be
the most convenient for multiphoton processes.

The analytical adiabatic approach is valid for {\em multiphoton} detachment
processes, i.e., strictly speaking, for $n\gg 1$. However, the calculations
for H$^-$ and halogen negative ions indicate (Gribakin and Kuchiev 1997a,
Kuchiev and Ostrovsky 1998) that the analytical formulae should
give reasonable answers even for $n=2$. 
To verify these conclusions we performed direct numerical calculations of the
two-photon detachment cross section of fluorine F$^-$ negative ion 
(Gribakin \etal 1998) within the lowest order of perturbation
theory and compared the results obtained with different ground-state
wavefunctions. We demonstrated explicitly the sensitivity of the cross
sections to the asymptotic behaviour of the bound-state wave function and
showed that the true cross section should be substantially higher than it was
previously believed, based on calculations with the HF $2p$ wavefunction.
Moreover, the use of the ground-state wavefunction with correct asymptotic
behaviour in multiphoton detachment calculations is often more important than
other effects of electron correlations.

In our previous work the asymptotically correct $2p$ wavefunction was
obtained using a model potential chosen to reproduce the experimental value
of the $2p$-electron energy. Of course, any model-potential approach is not
free from ambiguities related to the choice of the potential. The aim of
present work is to perform more accurate calculations of the two-photon
detachment in negative halogen ions using correct $np$ wavefunctions obtained
within the many-body Dyson equation method (see, e.g.
Chernysheva \etal 1988). Section 2 briefly outlines the method of calculation.
A discussion of our results and comparisons with other calculations and
experimental data are presented in Section 3.

\section{Method of calculation}\label{Meth}

\subsection{Two-photon detachment}

The total cross section of two-photon detachment of an electron from an
atomic system by a linearly polarized light of frequency $\omega $ is
\begin{equation}\label{cr}
\sigma =\sum _{l_fL} \sigma _{l_fL}=\frac{16\pi ^3}{c^2}\omega ^2
\sum _{l_fL}\left| A_{l_fL}( \omega ) \right| ^2~,
\end{equation}
where $\sigma _{l_fL}$ is the partial cross section for the final-state
photoelectron orbital momentum $l_f$ and total orbital momentum $L$, and the
continuous-spectrum wavefunction of the photoelectron is normalized to the
$\delta $-function of energy. For the detachment of the outer $np$ electron
from the negative halogen ion $np^6~^1S$ the final state can be
either $^1S$ ($L=0$, $l_f=1$) or $^1D$ ($L=2$, $l_f=1,~3$). In the lowest
2nd order the two-photon amplitude $A_{l_fL}( \omega )$ is determined by the
following equations
\begin{equation}\label{me1}
A_{l_fL}( \omega ) =\sqrt{2L+1}\left( 
\begin{array}{ccc}
1 & 1 & L \\
0 & 0 & 0
\end{array}
\right)
\sum _l (-1)^l \left\{ 
\begin{array}{ccc}
1 & 1 & L \\ 
l_f & l_0 & l
\end{array}
\right\} M_{l_fl}( \omega )~,
\end{equation}
\begin{equation}\label{me2}
M_{l_fl}( \omega ) =\sum _\nu \frac{\langle \varepsilon _fl_f \|\hat d \|
\nu l \rangle \langle \nu l\| \hat d \| n_0l_0\rangle }
{E_0 +\omega -E_\nu + \i 0 }~,
\end{equation}
where $\nu l$ is the intermediate electron state with the orbital momentum
$l$ after absorbing the first photon ($l=0,2$ for the halogens), and $n_0l_0$
is the initial bound state. The reduced dipole matrix elements are defined
in the usual way, e.g., in the length form,
\begin{equation}\label{dme3}
\langle \nu l\| \hat d \| n_0l_0\rangle =(-1)^{l_>}\sqrt{l_>}
\int P_{\nu l}(r) P_{n_0l_0}(r) r\d r ,
\end{equation}
where $l_>=\mbox{max}\{ l,l_0\}$ and $P$'s are the radial wave
functions.

The photoelectron angular distribution is given by the differential cross
section
\begin{equation}\label{dcr}
\frac{d\sigma }{d\Omega }=\frac \sigma {4\pi }\sum _{j=0}^2
\beta_{2j}( \omega )P_{2j}( \cos \theta ) ~,
\end{equation}
where $\theta $ is measured with respect to the light polarization axis,
and the asymmetry parameters $\beta _{2j}$ are determined in
terms of the two-photon transition amplitudes $A_{l_fL}$ and scattering
phases of the photoelectron $\delta _{l_f}$:
\begin{eqnarray}\label{beta}
\fl \beta _{2j}=\frac{16\pi ^3\omega ^2}{c^2\sigma }(4j+1)\mbox{Re}
\Biggl[ \sum _{l_f^\prime L^\prime l_f^{\prime \prime } L^{\prime \prime }}
(-1)^{l_0+L^\prime +L^{\prime \prime}}
(-\i )^{l_f^\prime +l_f^{\prime \prime }}
\exp \left[\i (\delta _{l_f^\prime}
-\delta _{l_f^{\prime \prime }})\right]
\sqrt{[l_f^\prime ][L^\prime ][l_f^{\prime \prime}]
[L^{\prime \prime }]} \nonumber \\
\times 
\left(
\begin{array}{ccc}
l_f^\prime & 2j & l_f^{\prime \prime} \\
0 & 0 & 0
\end{array}
\right)
\left( 
\begin{array}{ccc}
L^\prime  & 2j & L^{\prime \prime } \\
0 & 0 & 0
\end{array}
\right)
\left\{ 
\begin{array}{ccc}
L^\prime & L^{\prime \prime } & 2j \\
l_f^{\prime \prime} & l_f^\prime & l_0
\end{array}
\right\}
A_{l_f^\prime L^\prime }A^*_{l_f^{\prime \prime} L^{\prime \prime }}
\Biggr]~,
\end{eqnarray}
where $[l]\equiv 2l+1$ and $\beta _0=1$, so that the photoelectron angular
distribution after two-photon detachment is characterized by
$\beta _2$ and $\beta _4$.

The wavefunctions of the intermediate ($\nu l$) and final
($\varepsilon _fl_f$) states of the photoelectron are calculated in the HF
field of the frozen neutral atom residue $np^5$. The photoelectron is coupled
to the core to form the total spin $S=0$ and angular momenta $L=1$
for the intermediate $s$ and $d$ states ($l=0,\,2$), $L=0,\,2$ for the final
state $p$ wave ($l_f=1$), and $L=2$ for the final-state $f$ wave ($l_f=3$).
The intermediate state continua are discretized and represented by a
70-state momentum mesh with constant spacing $\Delta k$.

\subsection{Ground-state wavefunction}\label{ground}

If one describes the initial state $n_0l_0$ in the HF approximation, the
asymptotic behaviour of the corresponding radial wavefunction is incorrect.
Namely, it is characterized by $\kappa $ corresponding to the HF binding
energy, rather than the exact (experimental) one. For example, in F$^-$ the
HF value is $\kappa =0.6$, whereas the true one is $\kappa =0.5$. As we have
shown (Gribakin \etal 1998) is in fact much more important to have
an asymptotically correct bound-state wavefucntion $P_{n_0l_0}$ than to use
correct initial state energy $E_0$ in equation (\ref{me2}). The need for 
an asymptotically correct wavefunction was also clearly illustrated by the
adiabatic hyperspherical calculation of multiphoton detachment from H$^-$ by
Liu \etal (1992), where a 3.4\% change of $\kappa $ resulted in a 25\% change
of the two-photon cross section. In our previous paper we corrected the $2p$
wavefunction by solving the HF equations for the F$^{-}$ ground state with
a small additional repulsive potential of the form
$V(r)=\alpha /[2(r^2+a^2)^2]$, where the parameters $\alpha$ and $a$ were
chosen to ensure that the calculated energy was equal to the experimental
value. Our choice $\alpha =1$ and $a=0.61$ ensured $\kappa =0.5$, and produced
the asymptotic parameter $A=0.86$, close to the value recommended by
Radtsig and Smirnov (1986).

In this work we refine the bound-state wavefunction using atomic many-body
theory methods. The latter enable one to obtain a quasi-particle orbital 
which describes the bound electron in a many-body system from the Dyson
equation (see, e.g., Chernysheva \etal 1988, Gribakin \etal 1990, where it is
applied to calculations of negative ions)
\begin{equation}\label{dem}
\hat H^{(0)}\phi_E({\bf r})+\int \Sigma _E({\bf r},
{\bf r}') 
\phi_E ({\bf r}')\d {\bf r}^\prime =E \phi_E ({\bf r}) 
\end{equation}
Here $H^{(0)}$ is the single-particle HF Hamiltonian and
$\Sigma_E$ is the self-energy of the single-particle Green's function.
This energy-dependent non-local operator plays the role of a correlation
potential, and, if known exactly, produces exact bound-state energies
from equation (\ref{dem}). In most applications $\Sigma_E$ is calculated
by means of perturbation-theory expansion, and the lowest second-order
contribution usually gives a considerable improvement on the HF results.
Note that in some sense the Dyson equation provides the best single-particle
orbital of the initial state for photodetachment calculations. Because it is
in fact a quasiparticle orbital, it incorporates certain many-body effects,
namely, the ground-state correlations leading to the correct binding energy.

Within the second order in the electron Coulomb interaction $V$ the
matrix element of $\Sigma_E$ between some single-electron states $a$ and $b$
looks like
\begin{eqnarray}\label{sigma}
\langle a|\Sigma _E|b\rangle &=&\sum _{\nu _1,\nu _2,n_1}
\frac {\langle an_1|V|\nu _1\nu _2\rangle (\langle \nu _2\nu _1|V|n_1b\rangle
- \langle \nu _1\nu _2|V|n_1b\rangle )}
{E-E_{\nu _1}-E_{\nu _2}+E_{n_1}+\i 0} \nonumber \\
&+& \sum _{\nu _1,n_1,n_2}
\frac {\langle a\nu _1|V|n_1n_2\rangle (\langle n_2n_1|V|\nu _1b\rangle
- \langle n_1n_2|V|\nu _1b\rangle )}
{E-E_{n_1}-E_{n_2}+E_{\nu _1}-\i 0}~,
\end{eqnarray}
where the sums run over occupied states $n_1$ and $n_2$, and 
excited states $\nu _1$ and $\nu _2$, and the second terms in parentheses are
exchange contributions. The lowest-order correction to energy of the orbital
$a$ is given by $\langle a|\Sigma _E|a\rangle $ calculated with $E=E_a$.

For all halogen negative ions the HF binding energies of the outer $np$
subshell are greater than the corresponding experimental electron affinities.
Hence, the correlation correction to the energy must be positive, which
means a repulsive correlation potential.
Indeed, our numerical calculations show that $\Sigma _E$ is dominated by
the direct contribution in the second sum in equation (\ref{sigma})
(first term in parenthesis). For $E\approx E_{np}$ the sum over the occupied
states is basically given by $n_1=n_2=np$ (in many-body theory language this
means that both holes are the outer $np$ subshell). It is obvious then that
this contribution has $E_{\nu _1}-E_{np}>0$ in the denominator, and with a
squared matrix element (for $a=b=np$) in the numerator, it is explicitly
positive.

It also follows from our calculations that $\Sigma _E$ calculated in the
second-order approximation overestimates the correlation correction. Therefore,
to obtain best ground-state orbitals for multiphoton detachment
calculations we introduce a free parameter $\eta $ before the (dominant) direct
term in the second sum of equation (\ref{sigma}). This parameter is then
chosen to reproduce experimental values of the $np$ energies from the Dyson
equation. Moreover, using different values of $\eta $ we can effectively
simulate the fine-structure splitting, and obtain the wavefunctions for both
fine-structure components of the $np^6$ subshell, the upper $np_{3/2}$ and
lower $np_{1/2}$, corresponding to the different binding energies. For heavier
halogen negative ions (Br and I) the splitting between them becomes quite
significant, and we account for different asymptotic behaviour of the
corresponding radial wavefunctions in calculations of multiphoton processes.
Since we use $\Sigma _E$ from equation (\ref{sigma}) in this semiempirical
way, only the contributions of dominant monopole and dipole atomic excitations
are included in the sums.

Note that the importance of large distances in multiphoton problems
(Gribakin and Kuchiev 1997a, 1997b) supports the use of the length form of the
photon dipole operator. This is in agreement with the results of Pan \etal
(1990) who showed that the two-photon detachment cross sections obtained with
the dipole operator in the velocity form are much more sensitive to the shift
of the photodetachment threshold and correlation corrections, while the length
form results are much more robust.
 
The two-photon amplitudes $M_{{l_f}l}$ (\ref{me2}) are calculated by direct
summation over the intermediate states. It involves accurate evaluation of
the free-free dipole matrix elements, and special attention is paid to
pole- and $\delta $-type singularities of the integrand (Korol 1994, 1997).

\section{Results}\label{Disc}

\subsection{Fluorine}

The main difference between the present calculation and our previous
work (Gribakin \etal 1998) is in the initial state $2p$ wavefunction.
The self-consistent HF calculation of the F$^{-}$ ground state yields the
$2p$-electron energy $E_{2p}^{\rm HF}=-0.362$ Ryd, much lower than
its true value equal to the negative of the experimental electron affinity
of F: $E_{2p}^{\rm exp }=-0.250$ Ryd (Hotop and Lineberger 1986). In the
previous work we used a model potential which reproduced the experimental
energy, and yielded a $2p$ wave function with the asymptotic parameters
$A=0.86$ and $\kappa =0.5$ (cf. $A=0.94$ and $\kappa =0.6$ for the HF
wavefunction).

In the present work we obtain the $2p$ wavefunction from the Dyson
equation (\ref{dem}). When $\Sigma_E$ is calculated within the second order,
equation (\ref{sigma}), the $2p$ energy equal to $-0.187$ Ryd is obtained.
Thus, the second-order approximation overestimates the strength of the
polarization potential. When a scaling factor $\eta =0.67$ is introduced
in the way outlined in Sec. \ref{ground}, we reproduce the experimental
energy for the $2p$ electron, and obtain an accurate Dyson orbital of
the $2p$ subshell. This $2p$ wavefunction is quite close to the HF one
inside the atom, whereas for $r>2$ au it goes higher than the HF solution.
Its asymptotic behaviour is characterized by $\kappa=0.5$ and $A=0.64$.
The latter value of $A$ together with our best asymptotic parameters of the
$np$ orbitals of the other ions are presented in Table \ref{A}, where they
are compared with values recommended by Radtsig and Smirnov (1986) and
Nikitin and Smirnov (1988). The latter were obtained by matching the HF
wave function with that possessing a correct asymptotic behaviour. Unlike
the use of the Dyson equation with the many-body correlation potential
(even if somewhat adjusted), this procedure is not free from ambiguities.
They manifest in the differences between $A$ values from the two sources
cited.

\begin{table}
\caption{Asymptotic parameters of the $np$ orbitals of the halogen negative
ions.}\label{A}
\begin{indented}
\item[]\begin{tabular}{clllll}
\br
Ion & Orbital & $\kappa ^{\rm a}$ & $A^{\rm b}$ & $A^{\rm c}$ & $A^{\rm d}$\\
\mr
F$^-$  & $2p^{\rm e}$ & 0.500 & 0.64  & 0.84 & 0.7 \\
Cl$^-$ & $3p^{\rm e}$ & 0.516 & 1.355 & 1.34 & 1.3 \\
Br$^-$ & $4p_{3/2}$   & 0.497 & 1.53  & 1.49 & 1.4 \\
Br$^-$ & $4p_{1/2}$   & 0.530 & 1.60  & $-$  & $-$ \\
I$^-$  & $5p_{3/2}$   & 0.474 & 1.808 & 1.9  & 1.8 \\
I$^-$  & $4p_{1/2}$   & 0.542 & 2.587 & $-$  & $-$ \\
\br
\end{tabular}
\item[$^{\rm a}$]~Obtained using experimental binding energies.
\item[$^{\rm b}$]~Obtained from the our solutions of the Dyson equation.
\item[$^{\rm c}$]~Radtsig and Smirnov (1986).
\item[$^{\rm d}$]~Nikitin and Smirnov (1988).
\item[$^{\rm e}$]~We neglect the fine-structure splitting for F$^-$ and Cl$^-$.

\end{indented}
\end{table}

\begin{figure}
\epsfxsize=13cm
\centering\leavevmode\epsfbox{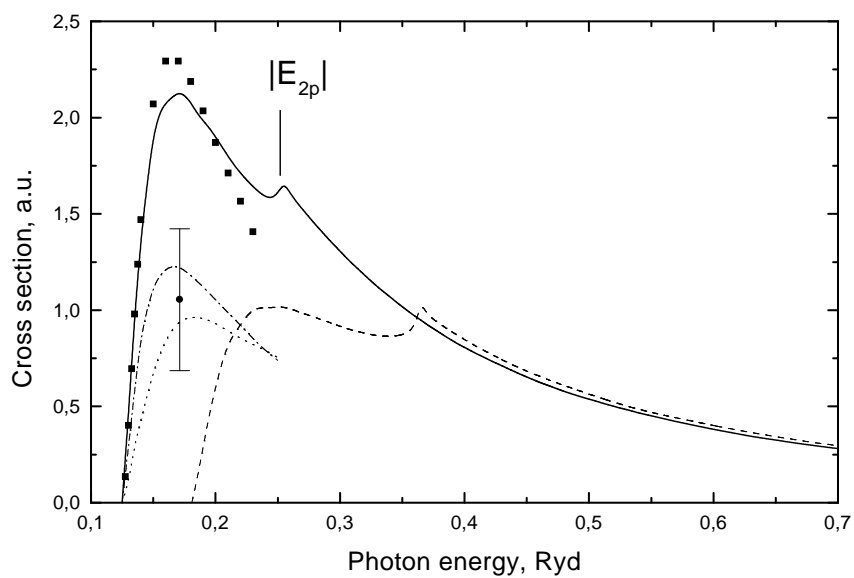}
\caption{Two-photon detachment cross sections of F$^-$. Present calculations:
\dashed , HF wavefunctions of the $2p$, intermediate and final states;
\full , same with the $2p$ wavefunction from the Dyson equation. Other
results: \dotted , calculations by Pan \etal (1990) with correlations 
and experimental binding energies; \chain , $R$-matrix Floquet method
by Hart (1996); $\protect\fullsqr $, model calculations by Robinson
and Geltman (1967); $\protect\fullcirc $,  experiment (Kwon \etal 1989). 
\label{fcross}}
\end{figure}

\begin{figure}
\epsfxsize=13cm
\centering\leavevmode\epsfbox{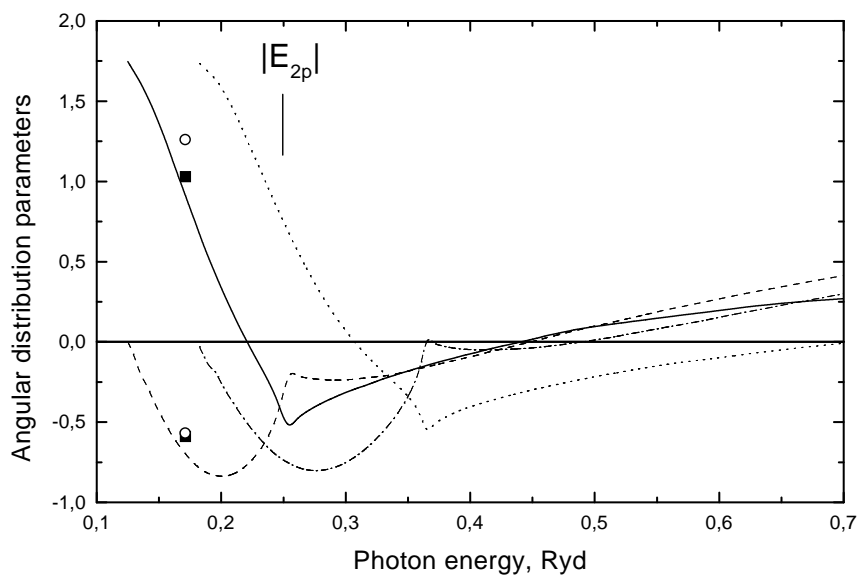}
\caption{Photoelectron angular distribution parameters for F$^-$.
\dotted, the HF $\beta_2 $ parameter; \chain , the HF $\beta_4 $ parameter;
\full , $\beta_2 $ parameter with the Dyson $2p$ wavefunction;
\dashed , $\beta_4 $ parameter with the Dyson $2p$ wavefunction.
$\protect\fullsqr $, experiment (Blondel and Delsart 1993);
$\protect\opencirc $, correlated length results by Pan and Starace (1991).
\label{fangul1}}
\end{figure}

The results of calculations of the two-photon detachment cross section and 
photoelectron angular distribution in F$^-$ are presented in figures 
\ref{fcross} and \ref{fangul1}. Our results obtained using the HF $2p$ orbital
(the HF two-photon threshold is at $\omega =0.181$ Ryd) are about
10\% higher than the similar dipole length lowest-order HF results 
of Pan \etal (1990). A possible source of this discrepancy was 
discussed in our previous paper (Gribakin \etal 1998), where we proposed
that it could be associated with the fact that Pan \etal (1990) used the
Roothaan-HF expansion of the bound state. According to them electron
correlation effects suppress the cross section in F$^-$ by about 20\% at the
maximum. 

When we use the experimental energy of the $2p$ electron together with
the HF wavefunctions the magnitude of the two-photon cross section changes
very little, as seen earlier by Pan \etal (1990) for both HF and correlated
results (dotted line in figure \ref{fcross}).
The HF results of Crance (1987a) are close to the above. The cross
section of van der Hart (1996) obtained within the $R$-matrix 
Floquet approach is 30\% higher (dash-dotted line
in figure \ref{fcross}) with a maximum of
$\sigma =1.25$ au at $\omega =0.166$ Ryd.

However, when we use the $2p$ wavefunction with the correct asymptotic
behaviour from the Dyson equation, the photodetachment cross section increases
about two times.\footnote{With the model $2p$ wavefunction, which had a larger
value of $A$, the cross section was about three times larger than the HF
results (Gribakin \etal 1998). As follows from the adiabatic theory
(Gribakin and Kuchiev 1997a, 1997b), for a given $\kappa $ the $n$-photon
cross section is proportional to $A^2$.}.
This cross section is shown by solid line in figure \ref{fcross}, and we
consider this to be our best evaluation of the cross section for F$^-$. 
The cusp on the curve at the single-photon threshold is a Wigner threshold
effect. It is a consequence of the abrupt threshold dependence
$\sigma \propto \sqrt{\omega -|E_0|}$ of the $s$-wave single-photon detachment
channel, which opens at this energy.

The only other work that used an asymptotically correct $2p$ wavefunction was
the model potential calculation of Robinson and Geltman (1967), which produced
a cross section two times greater than those of Crance, Pan \etal and
van der Hart. The results of Robinson and Geltman (full squares in figure
\ref{fcross}) are much closer to our results in comparison with all others.

Thus, we see that in {\em multiphoton} processes the error introduced by
using a bound-state wavefunction with an incorrect asymptotic behaviour
could be much greater then the effects of electron correlations.
Of course, the difference between the experimental and HF energies is also
a manifestation of electron correlation effects. It influences the result via
the asymptotic behaviour of the ground-state wavefunction, and we see that
this is the most important correlation effect in multiphoton detachment.
The use of the asymptotically correct $2p$ wavefunction changes the
cross section by a factor of two and more, which is much greater than other
correlation effects (Pan \etal 1990). This fact distinguishes this problem
from the single-photon processes, where other correlation effects are
essential.

The angular asymmetry parameters $\beta _2$ and $\beta _4$ calculated
using the Dyson $2p$-state wavefunction are shown in figure \ref{fangul1}.
They reveal an interesting dependence on the photon energy with sign
changes and cusps at the single-photon detachment threshold.
Figure \ref{fangul1} also presents the experimental points of Blondel and
Delsart (1993) obtained at $\omega =0.171$ Ryd, and the correlated dipole
length results of Pan and Starace (1991) at the same energy. The asymmetry
parameters (\ref{beta}) are relative quantities, and the results of different
calculations are much closer for them than for the absolute values of the
photodetachment cross sections. Thus, our present results are practically
equal to those obtained in the model $2p$ wavefunction calculation
(Gribakin \etal 1998). There we also showed that the results from
the analytical adiabatic theory (Gribakin and Kuchiev 1997b)
are in good agreement with the numerical results from other approaches,
including the plane-wave approximation, especially in $\beta _4$. It appears
that this parameter is on the whole less sensitive to the details of the
calculation, because it is simply proportional to the amplitude of $f$ wave
emission, and there are no interference terms in expression (\ref{beta}) for 
$\beta _4$. The experimental values of $\beta _{2,4}$ for F$^-$ obtained in
the earlier work of Blondel \etal (1992) are close to those of Blondel and
Delsart (1993). This is why F$^-$ serves as a good benchmark for angular
asymmetry calculations. Figure \ref{fangul1} shows that our present
calculations with the Dyson $2p$ wavefunction is in good agreement with
experiment.

\subsection{Chlorine}

The HF calculations of the Cl$^{-}$ ground state gives $E_{3p}^{\rm HF}=-0.301$
Ryd for the $3p$-electron energy, while the experimental value derived
from the electron affinity of Cl is $E_{3p}^{\rm exp }=-0.2657$ Ryd (Hotop
and Lineberger 1985). The spin-orbit splitting in Cl$^-$ is still small,
since the energies of the $3p_{3/2}$ and $3p_{1/2}$ states differ by
less than $0.008$ Ryd (Radtsig and Smirnov 1986), and we ignore it here.

When $\Sigma_E$ includes the monopole and dipole terms, equation (\ref{dem})
yields the $3p$ binding energy of $0.22$ Ryd. To obtain the $3p$ wavefunction
with the experimental $3p$ energy we solve equation (\ref{dem}) using
$\Sigma_E$ with the scaling factor $\eta =0.842$, introduced as outlined in
Sec. \ref{ground}. The corresponding asymptotic parameters are listed in
Table \ref{A}.

\begin{figure}
\epsfxsize=13cm
\centering\leavevmode\epsfbox{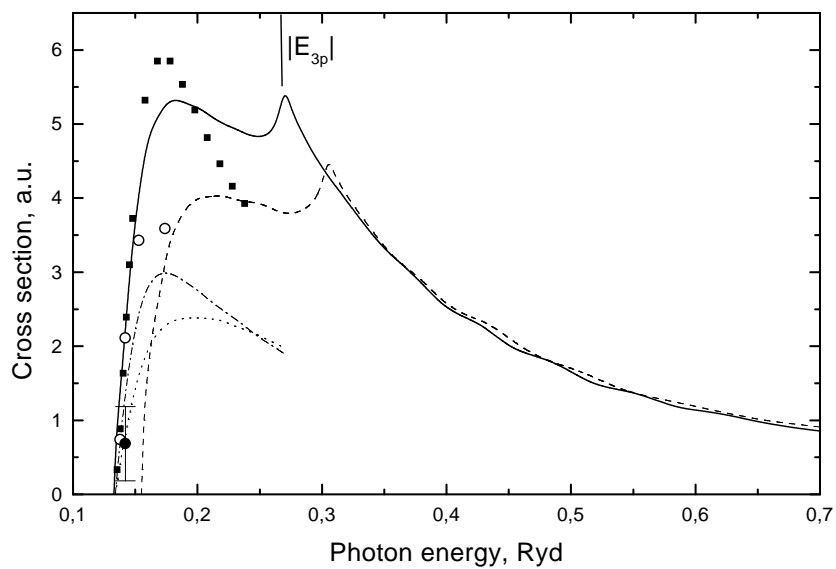}
\caption{Two-photon detachment cross sections of Cl$^-$. Present calculations:
\dashed , HF wavefunctions of the $3p$, intermediate and final states;
\full , same with the $3p$ wavefunction from the Dyson equation.
Other results: \dotted , calculations by Jiang and Starace (1988); 
\chain , $R$-matrix Floquet method by Hart (1996); 
$\protect\opencirc $, HF calculations by Crance (1987);
$\protect\fullsqr $, model calculations by Robinson
and Geltman (1967); $\protect\fullcirc $,  experiment (Trainham \etal 
1987).
\label{clcross}}
\end{figure}

The two-photon detachment cross sections we have obtained for the negative
chlorine ion are plotted in figure \ref{clcross} together with the results of
other calculations (Robinson and Geltman 1967, Crance 1987b, Jiang and Starace
1988, van der Hart 1996). Note that our HF cross section is significantly
higher than the results obtained by Jiang and Starace (1988)  and 
van der Hart (1996) (dotted and dotted-dashed lines in figure 
\ref{clcross}, respectively), but agree better with the HF cross section
calculated by Crance (1987) (open circles) and less than the cross sections 
obtained within plane-wave approximation (Crance 1987, Sturrus \etal 1992).

The total detachment cross section calculated with the Dyson $3p$ wavefunction
is 30\% higher at the maximum than the HF one, but above the single-electron
threshold it agrees quite well with the HF results. The maximal value of
the cross section is closer to the model calculations by Robinson and Geltman
(1967) (full squares in figure \ref{clcross}), who used an asymptotically
correct wavefunction from a model-potential calculation. However, the latter
calculation reveals a rapid decrease of the cross section beyond the maximum.
The experimental result obtained by Trainham \etal (1987) at $\omega =0.142$
Ryd is also shown.

\begin{figure}
\epsfxsize=13cm
\centering\leavevmode\epsfbox{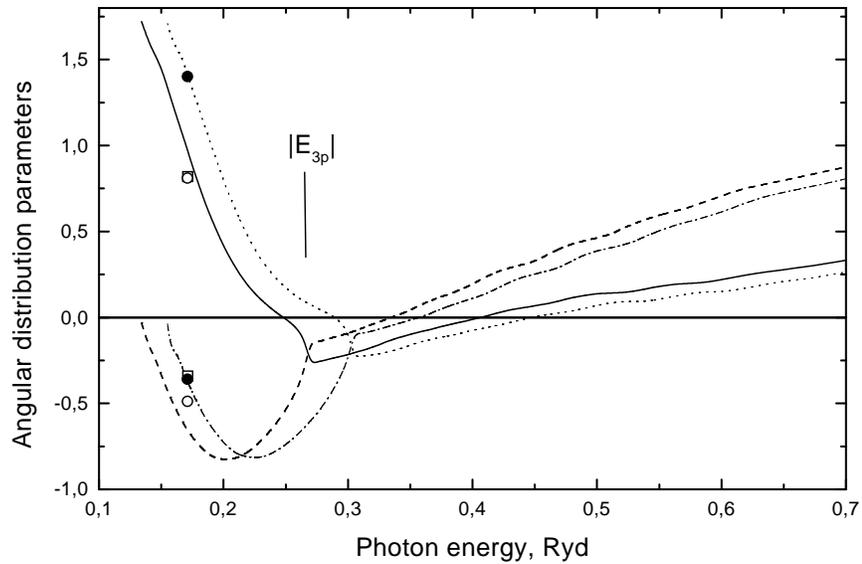}
\caption{Photoelectron angular distribution parameters for Cl$^-$.
\dotted, the HF $\beta_2 $ parameter; \chain , the HF $\beta_4 $ parameter;
\full , $\beta_2 $ parameter with the Dyson $2p$ wavefunction;
\dashed , $\beta_4 $ parameter with the Dyson $2p$ wavefunction.
$\protect\opensqr $, experiment (Blondel \etal 1992); 
$\protect\fullcirc $ and 
$\protect\opencirc $, calculations in the plane-wave approximation 
and with the first Born correction, respectively (Blondel \etal 1992).
\label{clangul}}
\end{figure}

The angular distribution parameters are shown in figure \ref{clangul}.
In general the parameters $\beta _2$ and $\beta _4$ behave similarly to
those in the negative fluorine, going through sign changes and cusps at the
single-electron threshold. Also presented are the experimental data and the
results of two calculations done within the plane-wave approximation and with
the first Born correction at $\omega =0.171$ Ryd (Blondel \etal 1992).
The first Born results are closer to our calculations with the correct
(Dyson) ground-state wavefunction, where the interaction between the
photoelectron and the atomic residue is included in all orders through the
HF wavefunctions of the photoelectron. Comparison with the experimental
data of Blondel \etal (1992) looks inconclusive. In the subsequent measurement
Blondel and Delsart (1993) confirmed the accuracy of the original data
for F$^-$, but found that the $\beta $ values for I$^-$ in Blondel \etal 1992
where affected by a spurious background. It appears that in the case of heavier
halogens, for the data obviously are not as reliable as in the fluorine case
(Blondel 1997).

\subsection{Bromine}

The binding energy of the $4p$ state in Br$^-$ obtained HF approximation is
equal to $0.2787$ Ryd, which is less than those for F$^-$ and Cl$^-$. On
the other hand, the spin-orbit splitting of the $4p$ orbital for Br$^{-}$,
0.034 Ryd (Radtsig and Smirnov 1986), is four times greater than that in the
negative chlorine ion. Besides this, the experimental data on the two-photon
detachment from Br$^-$ (Blondel \etal 1992) were obtained separately for the
two final fine-structure states of the atom, $4p_{3/2}$ and $4p_{1/2}$.
Therefore, we use different ground state wavefunctions for the
$4p_{3/2}$ and $4p_{1/2}$ states.

To obtain the $4p_{3/2}$ radial wavefunction we solve equation (\ref{dem})
with $\eta =0.865$ in $\Sigma _E$, and obtain the eigenvalue $E_{4p}=-0.247$
Ryd, equal to the experimental energy of the $4p_{3/2}$ state
(Hotop and Lineberger 1985). The experimental energy of the $4p_{1/2}$ state,
$-0.281 $ Ryd\footnote{It is equal to the negative of the sum of the electron
affinity and the fine-structure splitting of the $4p$ subshell in Br.}
is very close to the HF $4p$ energy, and we simply use the corresponding
HF wavefunction for the $4p_{1/2}$ orbital. The two wavefunctions thus
obtained have the following asymptotic parameters:
$A=1.53$ and $\kappa =0.497$ au for $4p_{3/2}$, and 
$A=1.60$ and $\kappa =0.530$ au for $4p_{1/2}$.
The wavefunctions of the intermediate and final states are calculated
in the frozen HF field of neutral atom residue, and we use the same sets to
calculate the two-photon amplitudes (\ref{me1}) and (\ref{me2})
for both $4p_{3/2}$ and $4p_{1/2}$ states.
After the electron detachment from these states the neutral Br is left in
either of the two fine-structure states $^2P_{3/2}$ or $^2P_{1/2}$. The
corresponding total and differential cross sections are evaluated from the
equations, which account for the number of electrons in the $j=\frac{3}{2}$
or $\frac{1}{2}$ sublevel of the $np$ subshell:
\begin{equation}\label{jcr}
\sigma ^{(j)}(\omega) =\frac{2j+1}{2(2l+1)} \sigma (\omega)
\end{equation}
\begin{equation}\label{jdcr}
\frac{d\sigma ^{(j)}}{d\Omega }=\frac{2j+1}{2(2l+1)}\frac
{d\sigma }{d\Omega}
\end{equation}
where $\sigma $ and $d\sigma /d\Omega $ are defined by equations
(\ref{cr}) and (\ref{dcr}), and $l=1$ for the halogen ions.

\begin{figure}
\epsfxsize=13cm
\centering\leavevmode\epsfbox{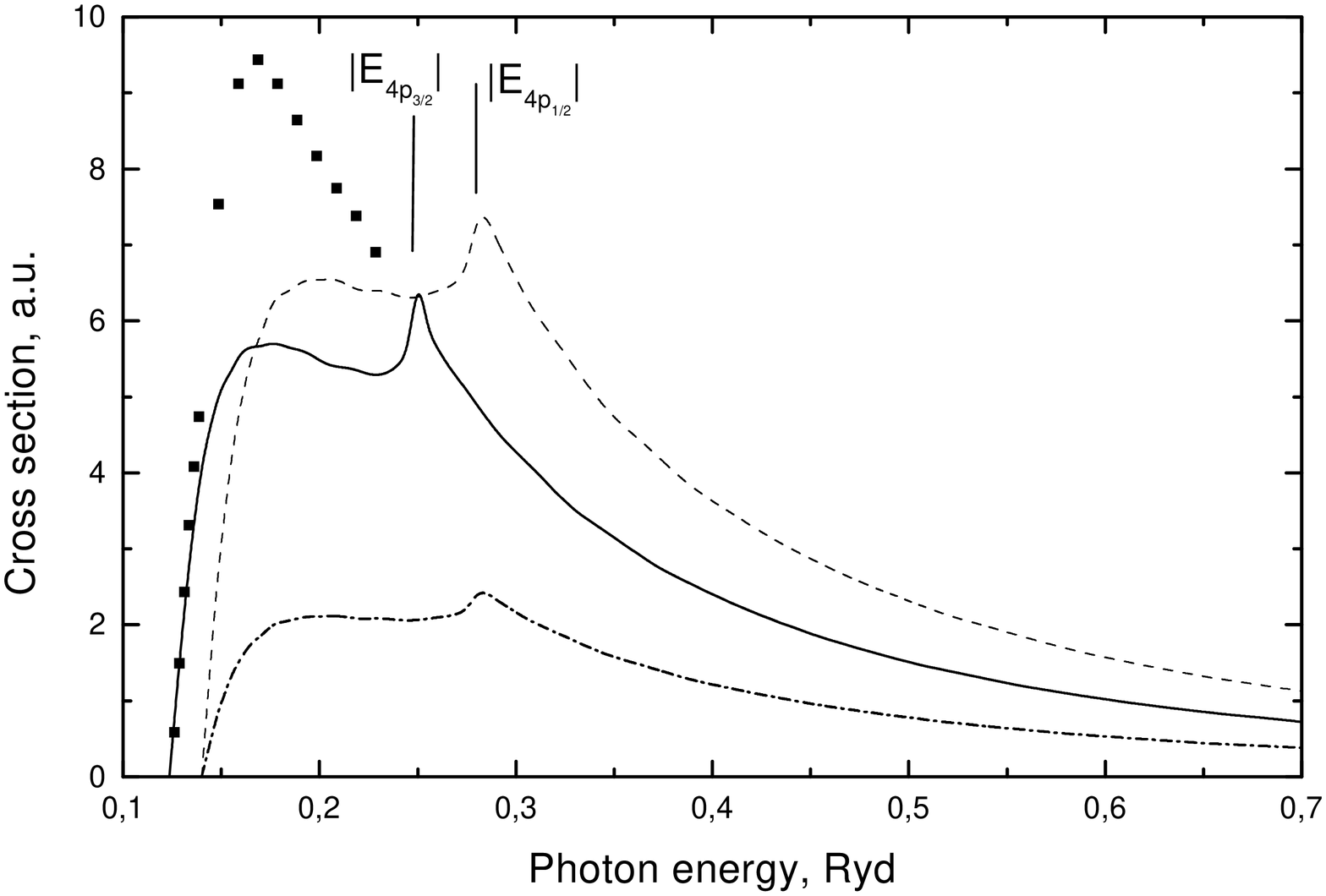}
\caption{Two-photon detachment cross sections of Br$^-$. 
\dashed , HF wavefunctions of the $4p$, intermediate and final states;
\full , same with the $4p_{3/2}$ wavefunction from the Dyson equation;
\chain ,  same with the correct $4p_{1/2}$ wavefunction; 
$\protect\fullsqr $, model calculations by Robinson and Geltman (1967).
\label{brcross}}
\end{figure}

\begin{figure}
\epsfxsize=13cm
\centering\leavevmode\epsfbox{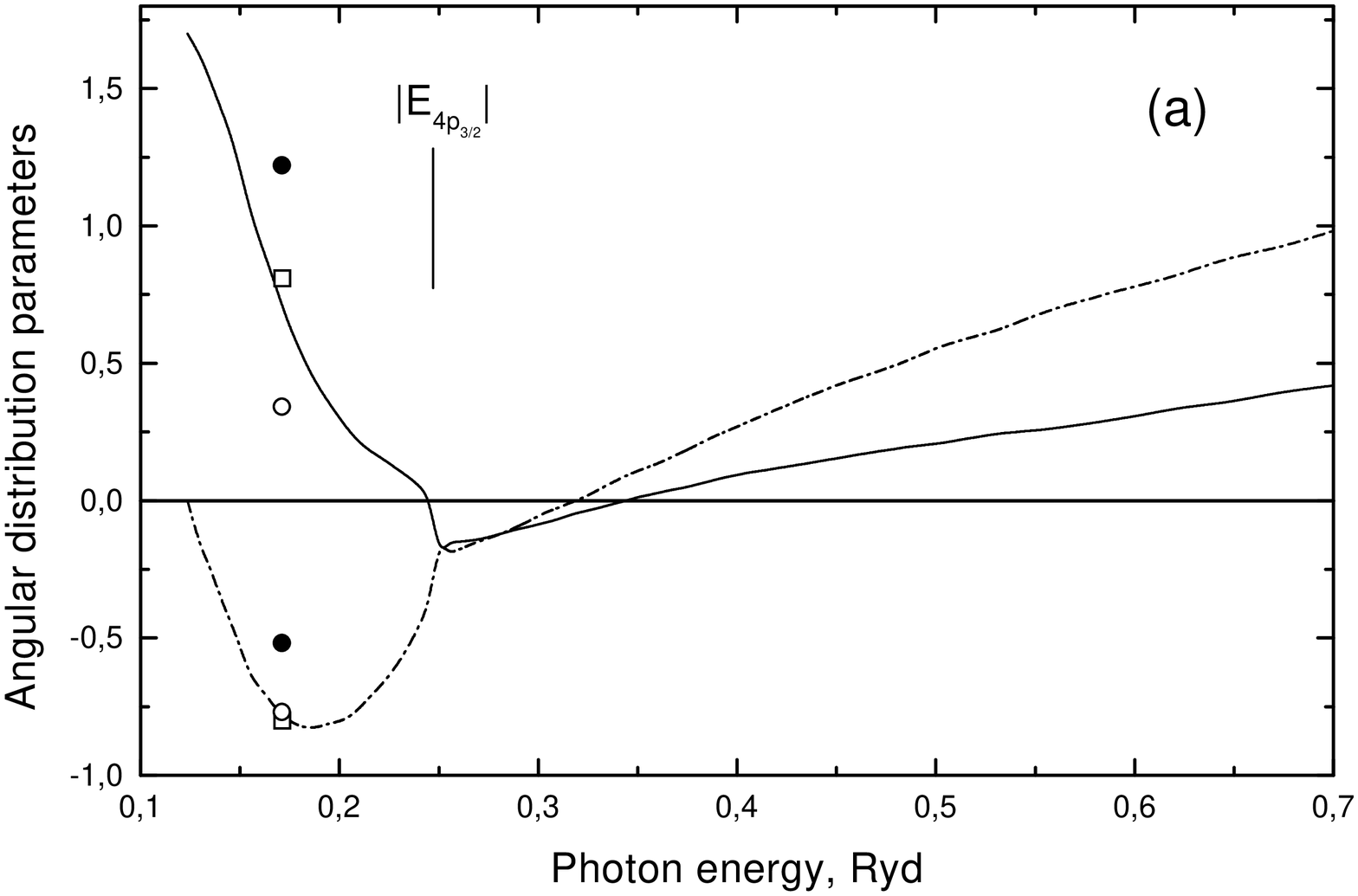}\\
\epsfxsize=13cm
\centering\leavevmode\epsfbox{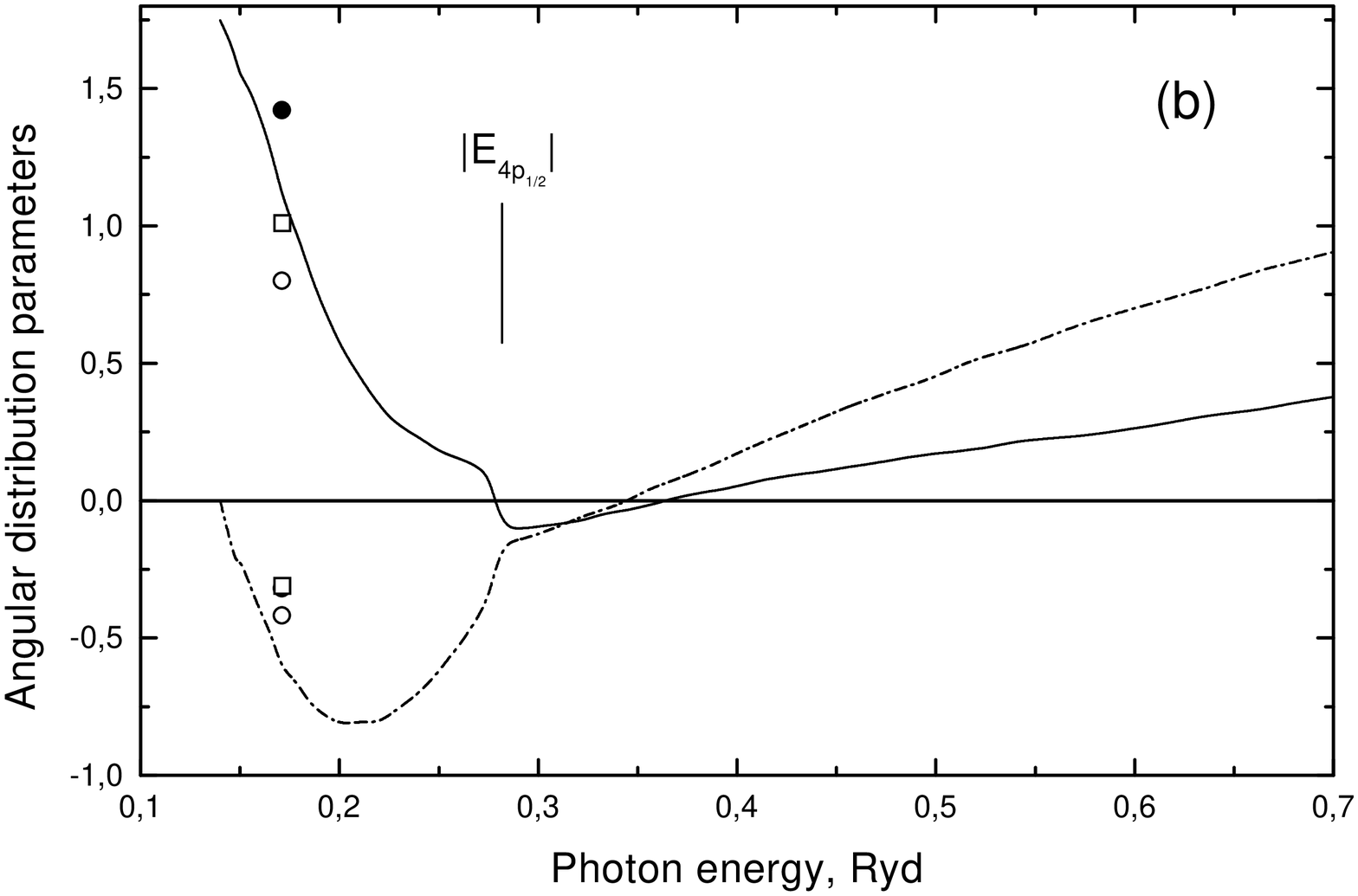}
\caption{Photoelectron angular distribution parameters for Br$^-$.
(a)  $\beta _2$ (\full) and $\beta _4$ (\dashed ) parameters 
for $4p_{3/2}$ state. $\protect\opensqr $, experiment (Blondel \etal 1992);
$\protect\fullcirc $ and $\protect\opencirc $, calculations 
in the plane-wave approximation and with the first Born correction, 
respectively (Blondel \etal 1992). 
(b) The same for the $\beta _2$ and $\beta _4$ parameters 
for $4p_{1/2}$ state.
\label{brangul}}
\end{figure}

The results of cross section and angular distribution parameter 
calculations are presented in figures \ref{brcross} and \ref{brangul}.

The partial $4p_{3/2}$ and $4p_{1/2}$ detachment cross sections have
similar shapes with the cusps shifted by the spin-orbit splitting energy.
We would like to emphasize that the ratio of the maxima of these partial
cross sections is not equal to the statistical $2:1$. The reason is that
because of the different binding energies the $4p_{3/2}$ and $4p_{1/2}$
wavefunctions have different asymptotic behaviour. As we discussed
above, this difference is enhanced in multiphoton processes, which clearly
favour the more loosely bound state.

We have not found any experimental values of the two-photon detachment
cross section in Br$^-$, so in figure \ref{brcross} we compare our results
with the calculations of Robinson and Geltman (1967) and Crance (1988).
In the calculation of Robinson and Geltman the fine-structure splitting of
was not taken into account. This means that they used the same wavefunction
and energy for both sublevels (basically, a $4p_{3/2}$ one, since it
corresponds to the experimental electron affinity). On the other hand, in the
HF calculation the wave function is practically identical to that of the
stronger bound $4p_{1/2}$ orbital. This is why our total cross section,
obtained as a sum $\sigma ^{(3/2)}+\sigma ^{(1/2)}$, goes between the
Robinson and Geltman and HF curves. The plane-wave approximation cross
section  of Crance (1988) has a much higher maximum than all of the results
presented in the figure.

As regards the photoelectron angular distribution, the asymmetry parameters 
for the $4p_{3/2}$ and $4p_{1/2}$ sublevels have similar dependences on the
photon energy (see figure \ref{brangul}). They also follow the general
trends observed for F$^-$ and Cl$^-$. The experimental data points
of Blondel \etal (1992) obtained at $\omega =0.171$ Ryd agree well with
our calculations, except for the $\beta _4$ parameter in the $4p_{1/2}$
detachment channel. Our calculation demonstrates better agreement with
experiment than the plane-wave approximation, or that which includes the
first Born correction (Blondel \etal 1992).

\subsection{Iodine}

The two-photon detachment calculation for I$^-$ are performed in the same
way as for Br$^-$. The experimental energies of the $5p_{3/2}$ and $5p_{1/2}$
fine-structure sublevels in I$^-$ are $-0.2248$ Ryd and $-0.2941$ Ryd,
respectively (Hotop and Lineberger 1985, Radtsig and Smirnov 1986).
The HF approximation yields $E_{5p}^{\rm HF}=-0.2583$ Ryd. The Dyson orbital
of the $5p_{3/2}$ state is obtained using the coefficient $\eta = 0.823$.
in $\Sigma_E$. It yields the experimental $5p_{3/2}$ state energy, and the 
wavefunction with asymptotic parameters $A=-1.808$ and $\kappa =0.474$ au.
The same calculations with $\eta = 0.008$ reproduces the experimental 
$5p_{1/2}$ energy, and produces a wavefunction with $A=-2.587$ and
$\kappa =0.542$ au. Using these ground-state wavefunctions together
with the HF sets of the intermediate and final state HF wavefunctions
we calculate the two-photon detachment amplitudes  (\ref{me1}) and (\ref{me2})
for the $5p_{3/2}$ and $5p_{1/2}$ electrons, and obtain the total and
differential cross sections (\ref{jcr}) and (\ref{jdcr}).

\begin{figure}
\epsfxsize=13cm
\centering\leavevmode\epsfbox{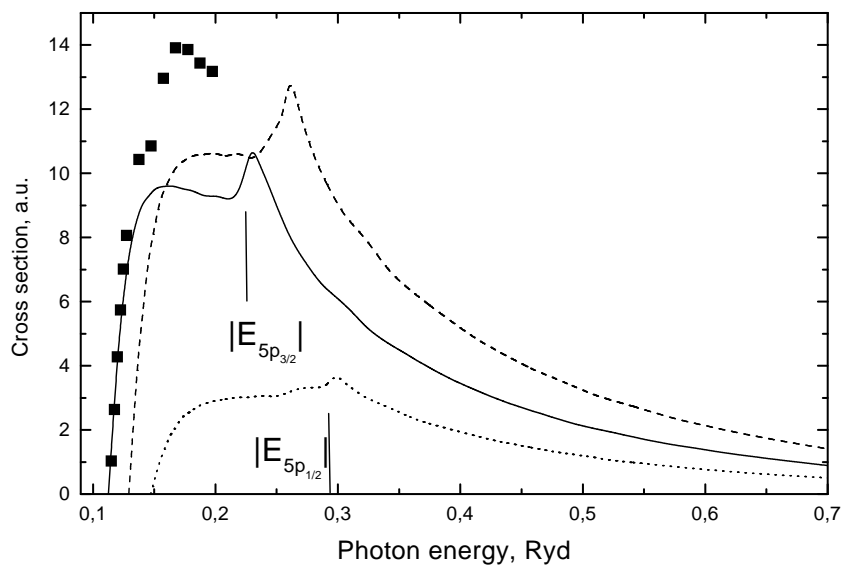}
\caption{Two-photon detachment cross sections of I$^-$. 
\dashed , HF wavefunctions of the $5p$, intermediate and final states;
\full , same with the $5p_{3/2}$ wavefunction from the Dyson equation;
\chain ,  same with the $5p_{1/2}$ wavefunction from the Dyson equation; 
$\protect\fullsqr $, model calculations by Robinson and Geltman (1967).
\label{icross}}
\end{figure}

The cross sections are shown in figure \ref{icross}. Due to a larger
difference in the binding energies the effect of non-statistical ratio
between the maxima of the $\sigma ^{(3/2)}$ and $\sigma ^{(1/2)}$ cross
sections is even more noticeable here than in Br. In I$^-$ their ratio in
the region below the single-photon threshold is about $3:1$ instead of the
statistical $2:1$. The HF threshold is
located between the two fine-structure thresholds. Nevertheless, the
sum $\sigma ^{(3/2)}+\sigma ^{(1/2)}$ would rise above the HF curve.
On the other hand, it would be close but still lower than the model-potential
results of Robinson and Geltman (1967). The plane-wave calculations
(Crance 1988) give an even higher cross section at the maximum.

\begin{figure}
\epsfxsize=13cm
\centering\leavevmode\epsfbox{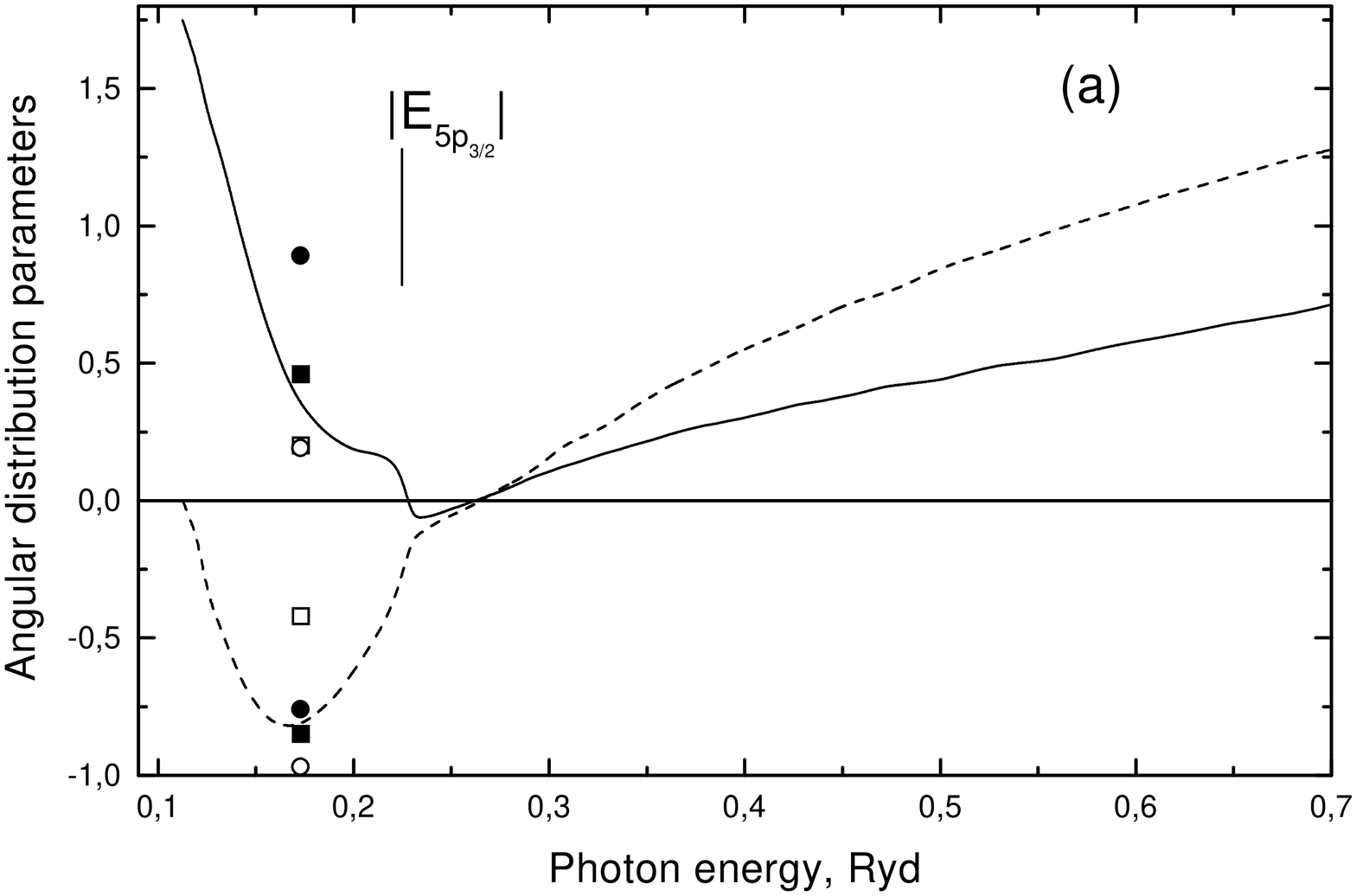}\\
\epsfxsize=13cm
\centering\leavevmode\epsfbox{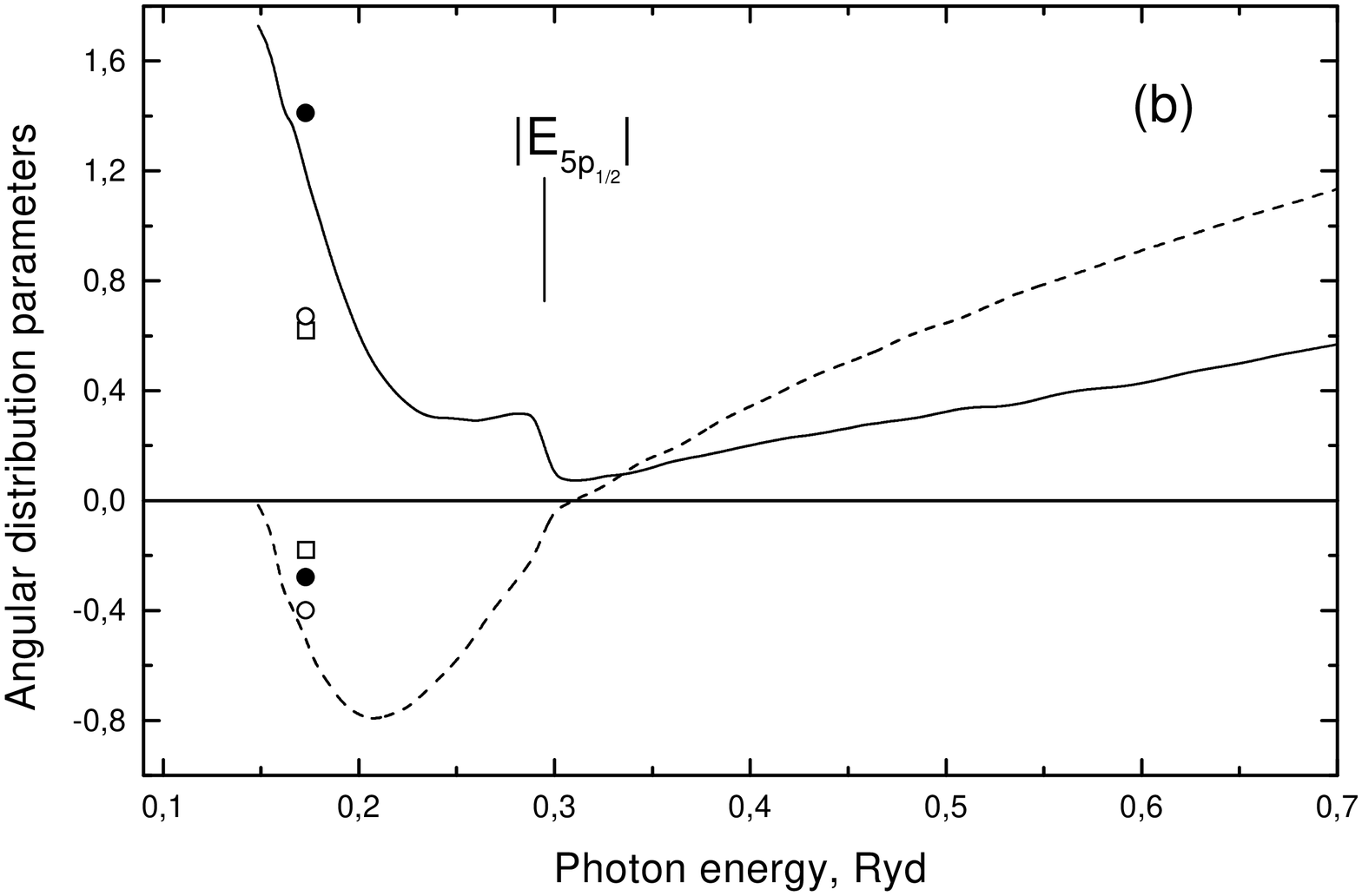}
\caption{Photoelectron angular distribution parameters for I$^-$.
(a)  $\beta _2$ (\full) and $\beta _4$ (\dashed ) parameters 
for $5p_{3/2}$ state. $\protect\opensqr $ and $\protect\fullsqr $, 
experimental data of Blondel \etal (1992) and Blondel and Delsart
(1993), respectively;
$\protect\fullcirc $ and $\protect\opencirc $, calculations
in the plane-wave approximation and with the first Born correction, 
respectively (Blondel \etal 1992). 
(b) The same for the $\beta _2$ and $\beta _4$ parameters 
for $5p_{1/2}$ state.
\label{iangul}}
\end{figure}

The angular distribution parameters are presented in figure \ref{iangul}.
For the $5p_{3/2}$ electron detachment from I$^-$ there are two measuments
at the photon energy of 0.171 Ryd (Blondel \etal 1992, Blondel and Delsart
1993), and the more recent one shows a much better agreement with our
calculations. If we speculate that the earlier measurement for the
$5p_{1/2}$ fine-structure component was affected by the spurious background in
a way similar to that seen in the $5p_{3/2}$ data, larger absolute values
of $\beta _{2,4}$ could be expected for the $5p_{1/2}$ detachment
measurements. This will bring them in better agreement with our calculated
values (figure \ref{iangul} b). As for the $\beta $ parameters calculated
in the plane-wave approach and with the first Born correction, they show large
scatter, similar to that seen in Cl and Br. It means that the potential
between the photoelectron and the atom should be included non-perturbatively,
at least at the HF level, as in the present calculation. Of course, for a more
accurate description one will have to include correlation effects, e.g., the
electron-atom polarization potential.

\section{Concluding remarks}\label{Concl}

In the present paper we have performed direct numerical calculations
of the two-photon detachment from the halogen negative ions. We paid
special attention to the proper description of the initial ground state
wavefunction, namely, its correct asymptotic behaviour. The outer $np$
ground-state orbitals of the negative ions were calculated from the many-body
theory Dyson equation with the non-local correlation potential adjusted to
reproduce experimental binding energies. We have confirmed the understanding
based on the adiabatic theory (Gribakin and Kuchiev 1997a, 1997b,
Gribakin \etal 1998) that using asymptotically correct initial state
wavefunctions is especially important for the absolute values of the
multiphoton detachment cross sections.

For heavier halogen negative ions (Br and I) our calculations reveal
substantial non-statistical branching of photodetachment into the
$P^2_{3/2}$ and $P^2_{1/2}$ final atomic states. This effect is mainly a
consequence of the different asymptotic behaviour of the corresponding
outer negative ion orbitals $np_{3/2}$ and $np_{1/2}$. Our calculations
also predict the existence of prominent cusps in the two-photon detachment
cross sections and angular asymmetry parameters at the single-photon
detachment thresholds. Our cross sections are in general closer to those
obtained by Robinson and Geltman (1967), who worked within a model-potential
approach and used asymptotically correct bound-state wavefunctions.
Our calculations of the photoelectron angular asymmetry parameters give best
overall agreement with the measurements of Blondel \etal (1992) and Blondel
and Delsart (1993) at the photon energy of 0.171 Ryd.

\section{Acknowledgements}\label{Ackn}

This work was supported by the Australian Research Council.
One of us (VKI) would like to acknowledge the hospitality extended to him
at the School of Physics at the University of New South Wales.

\section*{References}

\begin{harvard}
\item[] Blondel C, Cacciani P, Delsart C and Trainham R 1989 
{\it Phys. Rev. A} {\bf 40} 3698

\item[] Blondel C, Crance M, Delsart C and Giraud A 1992
{\it J. Physique II} {\bf 2} 839

\item[] Blondel C and Delsart C 1993 {\it Nucl. Instrum. Methods B}
{\bf 79} 156

\item[] Blondel C 1997 Private communication

\item[] Chernysheva L V, Gribakin G F, Ivanov V K and  Kuchiev M Yu 
1988 {\it J. Phys. B: At. Mol. Phys.} {\bf 21} L419-25

\item[] Crance M 1987a {\it J. Phys. B: At. Mol. Phys.} {\bf 20} L411

\item[]\dash 1987b {\it J. Phys. B: At. Mol. Phys.} {\bf 20} 6553-62

\item[] Crance M 1988 {\it J. Phys. B: At. Mol. Phys.} {\bf 21} 3559

\item[] Davidson M D, Broers B, Muller H G and van Linden van den Heuvell
1992 {\it J. Phys. B: At. Mol. Phys.} {\bf 25} 3093

\item[] Gribakin G F, Gul'tsev B V, Ivanov V K and Kuchiev M Yu 1990 
{\it J. Phys. B: At. Mol. Opt. Phys.} {\bf 23} 4505 

\item[] Gribakin G F and Kuchiev M Yu 1997a {\it Phys. Rev. A} {\bf 55}
3760

\item[] Gribakin G F and Kuchiev M Yu 1997b {\it J. Phys. B:
At. Mol. Opt. Phys.} {\bf 30} L657

\item[] Gribakin G F, Ivanov V K, Korol A V and Kuchiev M Yu 1998
{\it J. Phys. B: At. Mol. Opt. Phys.} {\bf 31} L589

\item[] Hall G L, Robinson E J and Branscomb L M 1965 {\it Phys.Rev.Lett.}
{\bf 14} 1013

\item[] Hotop H and Lineberger W C 1985 {\it J. Phys. Chem Ref. Data}
{\bf 14} 731

\item[] Jiang T-F and Starace A F 1988 {\it Phys. Rev. A} {\bf 38}
2347-55

\item[] Keldysh L V 1964 {\em Zh. Eksp. Teor. Fiz.} {\bf 47} 1945
[1965 {\em Sov. Phys. JETP} {\bf 20} 1307].

\item[] Korol A V 1994 {\it J. Phys. B: At. Mol. Phys.} {\bf 27} 155

\item[] Korol A V 1997 unpublished

\item[] Kuchiev M Yu and Ostrovsky V N 1998 {\it J. Phys. B:
At. Mol. Opt. Phys.} {\bf 31} 2525-38

\item[] Kwon N, Armstrong P S, Olsson T, Trainham R and Larson D J 1989
{\it Phys. Rev. A} {\bf 40} 676

\item[] Liu C-R, Gao B and Starace A F 1992 {\it Phys. Rev. A} {\bf 46}
5985

\item[] Nikitin E E and Smirnov B M 1988 {\it Atomic and Molecular
Processes} (Moscow: Nauka) p~283

\item[] Pan C, Gao B and Starace A F 1990 {\it Phys. Rev. A} {\bf 41} 6271

\item[] Pan C and Starace A F 1991 {\it Phys. Rev. A} {\bf 44} 324

\item[] Radtsig A A and Smirnov B M 1986 {\it Parameters of Atoms and
Atomic Ions} (Moscow: Energoatomizdat)

\item[] Robinson E J and Geltman S 1967 {\it Phys. Rev.} {\bf 153} 4

\item[] Sturrus W J, Ratliff L and Larson D J 1992 {\it J. Phys. B: 
At. Mol. Phys.} {\bf 25} L359

\item[] Trainham R, Fletcher G D and Larson D J 1987 {\it J. Phys. B: 
At. Mol. Phys.} {\bf 20} L777

\item[] van der Hart H W 1996 {\it J. Phys. B: At. Mol. Phys.} {\bf 29}
3059-74

\end{harvard}

\Figures

\end{document}